\documentclass[a4paper]{article}
\usepackage{graphicx} 
\usepackage{bbm}
\usepackage{times} 
\usepackage{float}
\usepackage{amsmath,amssymb,amsfonts}
\usepackage[latin1]{inputenc}
\usepackage[english]{babel}
\usepackage[cyr]{aeguill}
\usepackage{xcolor}
\usepackage{cite}
\usepackage{multicol}
\usepackage{multirow}
\usepackage{array}
\usepackage{booktabs}
\usepackage{appendix}
\usepackage{colortbl}

\usepackage[pdftex,hyperindex=true,hyperfigures=true,bookmarks=true,pdftitle={TITLE},pdfsubject={},pdfauthor={AUTHOR},pdfkeywords={KEYWORDS},pdfpagemode=UseOutlines,colorlinks=true,urlcolor=blue,citecolor=red,linkcolor=blue]{hyperref}

\usepackage{enumerate,tikz}
 \ifx\pdfoutput\undefined 
    \usepackage[dvips]{graphficx}
 \usepackage[]{graphicx}
\fi
\usepackage{flafter} 
\DeclareGraphicsExtensions{.jpg,.mps,.pdf,.png} 
\usepackage[sf]{caption2}
\usepackage[figuresright]{rotating}

\usepackage{amsthm,amsmath,amssymb,amsbsy,mathrsfs,supertabular,eurosym,graphicx,enumitem,xcolor}
\usepackage{mathtools}

\graphicspath{
{./Figures/}
}

\makeindex  
\makeglossary  

\newtheorem{prop}{Proposition}

\newcommand{\djk}{d_{j,k}}
 \def\RR{\mathbb{R}}

\newcommand{\Hmin}{{H_{min}}}

\newcommand{\EEE}{{\rm I\kern-2pt E}}
\newcommand{\PP}{{\rm I\kern-2pt P}}

\newcommand{\dws}{D^{ws}_f}

\newcommand{\ome}{\omega}

\newcommand{\BE}{\begin{equation}}
\newcommand{\EE}{\end{equation}}

\makeatletter
\newcounter {subsubsubsection}[subsubsection]
\renewcommand\thesubsubsubsection{\thesubsubsection .\@alph\c@subsubsubsection}
\newcommand\subsubsubsection{\@startsection{subsubsubsection}{4}{\z@}%
                                     {-3.25ex\@plus -1ex \@minus -.2ex}%
                                     {1.5ex \@plus .2ex}%
                                     {\normalfont\normalsize\bfseries}}
\newcommand*\l@subsubsubsection{\@dottedtocline{3}{10.0em}{4.1em}}
\newcommand*{\subsubsubsectionmark}[1]{}
\makeatother

\newsavebox{\fmbox}
\newenvironment{fmpage}[1]
 {\begin{lrbox}{\fmbox}\begin{minipage}{#1}}
 {\end{minipage}\end{lrbox}\fbox{\usebox{\fmbox}}}

\usepackage{geometry}
\normalsize

\newtheorem{theo}{Theorem} 
\newtheorem{coro}[prop]{Corollary} 
\newtheorem{lemm}[prop]{Lemma}
\newtheorem{defi}{Definition}

\newcommand{\BP}{\begin{prop}}
\newcommand{\EP}{\end{prop}}
\newcommand{\BL}{\begin{lemm}}
\newcommand{\EL}{\end{lemm}}
\newcommand{\BD}{\begin{defi}}
\newcommand{\ED}{\end{defi}}
\newcommand{\BT}{\begin{theo}}
\newcommand{\ET}{\end{theo}}
\newcommand{\BC}{\begin{coro}}
\newcommand{\EC}{\end{coro}}

\providecommand{\keywords}[1]
{
  \small	
  \textbf{\textit{Keywords---}} #1
}

\title{ Multifractal Analysis of Physiological Signals: A Novel Approach to Optimizing Pacing Strategy in a Pilot Study}
\author{
    Wejdene Ben Nasr
    \thanks{Univ Paris Est Creteil, Univ Gustave Eiffel CNRS, LAMA
        UMR8050, F-94010 Creteil, France}, 
    V\'eronique Billat$\; ^{\ast} \; ^{\ddag}$
    \thanks{ {Universit\'e Paris-Saclay, Univ Evry, F-91000 Evry-Courcouronnes, France}}, 
    St\'ephane Jaffard$^{\ast}$,  \\ 
        Florent Palacin\thanks{Laboratoire de neurophysiologie et de biom\'ecanique du mouvement, 
        Institut des neurosciences de l'Universit\'e Libre de Bruxelles, Belgique \\ {veronique.billat@billatraining.com, wejdene.nasr-ben-hadj-amor@u-pec.fr, jaffard@u-pec.fr, palacinflorent@gmail.com, guillaume.saes@u-pec.fr}}, Guillaume Sa\"es$^{\ast}$ 
}

\begin{document}

\maketitle

\abstract{ 
Marathons are one of the ultimate challenges of human endeavor. In this paper, we apply recently introduced multifractal techniques which yield a new classification parameter in the processing of physiological data captured on marathon runners. The comparison of their values gives a new insight on the way that runners of different level conduct their run, and ultimately, can be used in order to give advice on how to improve their performance.  
}


\section{Introduction}


Marathon running requires a balance between speed and stamina and analyzing physiological signals can provide insights into performance, pacing strategy, and fatigue. Therefore, in this study we test the hypothesis that multifractal techniques
\begin{itemize}
    \item provide new insights into how runners of different levels manage their runs, 
    \item offer advice on improving performance,
     \item help detect fatigue and optimize pacing strategies.
\end{itemize}
Some traditional methods  have been used to reveal hidden dynamics in the data: 
\begin{itemize}
    \item the analysis of Heart Rate Variability (HRV) to monitor the autonomic nervous system's response during marathons, analyzing the effects of fatigue and endurance on heart rate patterns \cite{ivanov1999},  
    \item Detrended Fluctuation Analysis (DFA)  which allows to identify scaling behavior in physiological signals like heart rate and speed, revealing the impacts of prolonged exercise and fatigue \cite{kantelhardt2002multifractal,wesfreid05a}, 
    \item multifractal analysis, which emerged as a powerful tool to qualitatively assess physiological signals, offering a more comprehensive perspective on individual marathon performance by offering new classification and model selection parameters based on scaling invariance exponents. 
    \end{itemize}
    
    Several variants of multifractal analysis  have been challenged on marathon runners'data. A first one, the Wavelet Transform Modulus Maxima method (WTMM) allows to estimate the multifractal spectrum (i.e. the fractional dimension of the singularity sets of a given order) of a signal \cite{muzy1991wavelets}.  It has been applied to heart rate time series, demonstrating that heart rate variability can be described by scaling laws, see \cite{wesfreid05a} where multifractal analysis is applied to the physiological signals of marathon runners, focusing on the detection of fatigue and the optimization of pacing strategies. Advanced multifractal techniques such as Multifractal Detrended Fluctuation Analysis (MF-DFA) and wavelet-based multifractal formalism were used to quantify the multifractality of time series data collected during marathon races \cite{billat2009detection}. They both  detect long-range correlations in non-stationary time series and quantify their multifractal properties. These studies have shown that multifractal techniques can effectively characterize the variability and complexity of physiological signals collected  on  marathon runners, offering new insights into endurance performance.

However, under extreme conditions, these methods of performance analysis often fall short of capturing the complex and dynamic nature of physiological responses because the signals do not satisfy the minimal regularity assumptions required for such an analysis.  That is the reason why, more recently, we have examined how cadence and stride variability during marathons can be analyzed using multifractal techniques based on alternative regularity exponents such as $p$-exponents (see Sec. \ref{sec:math} below).  In some cases, even this extended version of multifractal analysis cannot be performed, and the widest possible setting is supplied the framework supplied by the weak-scaling pointwise regularity exponent introduced by Yves Meyer \cite{MeyWVS}; indeed, it provides a robust framework for analyzing signals without assuming any a priori global regularity on the data; a first application of this alternative technique has been performed in \cite{JSWPB}, where the focus was put on cadence (in this study, the analysis of velocity was not performed due to GPS measurement inaccuracies). 

The key findings of the present work include the following results: 
\begin{itemize}
    \item Initial raw data showed significant variations during different phases of the marathon, including warm-ups and breaks.
    \item After cleaning the data to remove non-marathon activities, continuous reconnections were  performed  to maintain homogeneity; this is theoretically possible without altering the regularity properties of the data, and therefore their multifractal properties  if regularity exponents are everywhere below 1  which we show the be the case here (see the Appendix for a proof of this result and an extension to the non locally bounded setting). 
    \item We show that  most data cannot be mathematically modelled by locally bounded functions, see Sec. \ref{sec:math} below; this  indicates the need for a multifractal analysis  analysis based  on $p$-exponents   if the data can be modeled by an $L^p$ function; finally,  if the data are so irregular that they cannot be modeled by a function belonging to any of the $L^p$ spaces, then one can have recourse to the weak-scaling exponent, see Sec. \ref{sec:math2}, which requires no a priori assumption on the data an can be worked out for general Schwartz distributions.
    \item  The study provided insights into how multifractal parameters can detect changes in physiological signals due to fatigue, particularly around the 30th kilometer mark where perceived exertion significantly increased.
\end{itemize}

Multifractal analysis is a  mathematical approach that goes beyond traditional  methods, allowing for the characterization of complex, irregular, and dynamic patterns data set.  Its variants have been successfully tested on many probabilistic models used in applications, one example being 
the multifractal random walk; indeed, this model provides a framework for understanding and simulating the multifractal nature of various complex systems, including financial markets, geophysics, and other fields where datasets exhibit non-linear and irregular behavior. This stochastic process  illustrates how multifractal analysis can go beyond traditional linear methods to capture the intricacies of complex datasets \cite{Bacry2001,abry2009multifractal}. In the  1990s,  the multifractal structure of  wavelet based multifractal analysis, was explored,  providing significant insights into the theoretical underpinnings of multifractal analysis, see \cite{Jaffard2004}. This work layed the ground for the application of multifractal methods in various fields, emphasizing the ability to describe complex, irregular patterns in datasets; it  promoted a variant of  multifractal analysis based on wavelet coefficients as a powerful tool in both theoretical and applied mathematics. Indeed, this method has proven effective in various fields, including finance, meteorology and medicine, due to its ability to capture intricate variability and interdependencies within dataset; various studies across different fields showed that multifractal analysis is a sophisticated mathematical approach that extends beyond traditional linear methods, see e.g.  \cite{CRAS2019} and ref. therein.

\begin{enumerate}
    \item \textbf{Physiological Data:} Multifractal analysis has been extensively applied to physiological data, such as in the study of heart rate variability of marathon runners. This method has been shown to capture the complex dynamics of physiological signals, offering insights into performance and the underlying health conditions \cite{JSW2,ACDJ}.
    \item \textbf{Finance:} Multifractal detrended fluctuation analysis (MF-DFA) is commonly used to study financial time series that exhibit volatility clustering and other irregular behaviors. This approach helps in identifying intricate patterns in market data, which are crucial for risk management and developing trading strategies \cite{jianga2018multifractal}.
\item \textbf{Meteorology:} Multifractal analysis also finds applications in meteorology, particularly in studying the joint influence of climatic variables like temperature, humidity, and evapotranspiration. These studies reveal how multifractal patterns can describe the complex interdependencies and variability in atmospheric data \cite{ariza2019joint}.
\item \textbf{Medicine:} The potential of using multifractality as a new biomarker for cancer diagnosis has been put in evidence, see \cite{gerasimova2014wavelet}, and more recently \cite{huynh2023multifractality}   where it allowed to   improve the differentiation between normal and cancerous cells  using adaptive versus median threshold for image binarization.
\end{enumerate}
These references demonstrate the versatility and effectiveness of multifractal analysis in uncovering hidden patterns and dependencies in diverse datasets, thereby confirming its value across various scientific domains. 

In a series of studies initiated by V. Billat, the physiological responses and pacing strategies of marathon runners were  extensively studied, providing a foundation for understanding how different variables interact during a race. Furthermore, in \cite{Billat19} the authors verified the hypotheses that "Mass runners try to maintain a constant speed without succeeding" and "Marathoners run in an asymmetric way, and this turns out to be visible in the speed time series". 
The objective of this new study is to demonstrate the potential of employing multifractal analysis for enhancing the runner's feedback regarding the optimization of their pacing strategy. To this end, the performance of marathoners who have completed the same marathon is compared in order to minimize the impact of the profile race. Subsequently, for each runner, the first 21 km and the last 10 km of the marathon are compared. For that purpose, we used  multifractal parameters to enhance the degree of adequate pacing strategy, thereby enabling the runner to achieve the race while maintaining the degree of high pace variation that has been demonstrated to be associated with personal bests, irrespective of the level of the runner (recreational or elite) \cite{Billat19}.

\section{The subjects characteristics and their performance.}

\begin{table}[H]
\begin{center}
\resizebox{170mm}{!}{
 \begin{tabular}{||c||*{11}{m{2.2cm}|}|} 
 \hline
  & \textcolor{blue}{M1} & M2 & M3 & \textcolor{blue}{M4} & \textcolor{blue}{M5} & M6 \textbf{Women} & \textcolor{blue}{M7} & \textcolor{blue}{M8} & \textcolor{blue}{M9} & M10 \\ [0.5ex] 
 \hline\hline
 Time & \textcolor{blue}{4:07:06} & 3:45:37 & 3:05:07 & \textcolor{blue}{2:52:24} & \textcolor{blue}{2:47:50} & 4:06:19 & \textcolor{blue}{4:13:35} & \textcolor{blue}{4:09:04} & \textcolor{blue}{3:22:19} & 0.4399 \\
 \hline
Marathon & \textcolor{blue}{Paris} & Tokyo & Montpellier & \textcolor{blue}{Paris} & \textcolor{blue}{Paris} & Sully sur Loire & \textcolor{blue}{Paris} & \textcolor{blue}{Paris} & \textcolor{blue}{Paris} & La Rochelle \\
\hline
Rank & \textcolor{blue}{10} & 7 & 5 & \textcolor{blue}{1} & \textcolor{blue}{2} & 4 & \textcolor{blue}{8} & \textcolor{blue}{9} & \textcolor{blue}{6} & 3 \\
\hline
Age & \textcolor{blue}{44} & 41 & 37 & \textcolor{blue}{50}& \textcolor{blue}{48} & 55 & \textcolor{blue}{53} & \textcolor{blue}{48} & \textcolor{blue}{32} & 52 \\
\hline
Weight (kg) & \textcolor{blue}{79} & 72 & 79 & \textcolor{blue}{65} & \textcolor{blue}{67} & 53 & \textcolor{blue}{83} & \textcolor{blue}{78} & \textcolor{blue}{80} & 75 \\
\hline
Height (cm) & \textcolor{blue}{180} & 173 & 185 & \textcolor{blue}{174} & \textcolor{blue}{174}& 170 & \textcolor{blue}{178} & \textcolor{blue}{171} & \textcolor{blue}{181} & 180 \\
 \hline
 \end{tabular}}
\end{center}
\captionsetup{width=0.9\textwidth}
\caption{Characteristics and Performance Metrics of Marathon Runners, including Time Performance, Marathon Name, Age, Weight, and Height. Runner M6 is the only female participant, and runners M4 and M5 represent the same individual with two different performances. Rankings are based on the percentage of time relative to the record for each gender and age category. Runners highlighted in blue are those who participated in the Paris Marathon  }
\label{tab-info}
\end{table}
Table \ref{tab-info} provides information about the subjects' characteristics and their performance, offering context and perspective on the results obtained. It includes data on the time performance, marathon name, age, weight, and height of experimented marathon runners, where all participants are men except for M6. Additionally, runners M4 and M5 represent the same person but with two different performances. The rank is determined by calculating the percentage of time taken in the race relative to the record of each category to which the marathon runner belongs based on  gender and age (a lower percentage of the record corresponds to a higher ranking for the marathon runner).

\section{Multifractality application}

\subsection{Mathematical framework}

\label{sec:math} 

Multifractal analysis  deals with the analysis and classification of everywhere irregular signals.  Its purpose  is to obtain estimates on the fractional dimensions of the sets of points where a pointwise regularity exponent takes a given value $H$. These dimensions, considered as a function of $H$, are referred to as the { \em multifractal spectrum}. It follows that a prerequisite of the method is to determine which notion of pointwise regularity is relevant for a given signal. Indeed, on the mathematical side,  several possible definitions have been introduced, each one making sense only in a particular functional setting. For instance, the one which is most widely used is the  pointwise H\"older exponent which is defined as follows.

\BD
 Let $x_0\in\RR$ and $\alpha\geq 0$;  a locally bounded finction $f$ belongs to $C^\alpha (x_0)$ if  there exist a polynomial $P_{f,x_0}$ of degree less than $\alpha$ and $C,r>0$ such that
\BE  \label{eq:reghol} 
\forall x \in (x_0-r,x_0+r), \quad |f(x)-P_{f,x_0}(x-x_0)| \leq C |x-x_0|^{\alpha}.
\EE
The H\"older exponent of $f$ at $x_0$ is $h_f (x_0)=\sup\{\alpha \ : \ f\in C^{\alpha} (x_0)\}$.
\ED 

It is important to note that this definition makes sense only if the data are locally bounded. Indeed \eqref{eq:reghol} implies  that $f$ is bounded in a neighbourhood of $x_0$. Therefore, if one wishes to use this notion of pointwise regularity for the analysis of some data, one has first to determine if these data can be modelled by a locally bounded function. This can be done easily using the wavelet decomposition of $f$, which we now recall. Let $\psi(x)$ be a  {\em wavelet}, i.e. a  well localized, smooth oscillating function such that  the 
\[  \psi_{j,k}(x) : = 2^{j/2} \psi (2^j x - k), \qquad j \in \mathbf{Z},  k \in \mathbf{Z} \]  form an orthonormal basis of $L^2(\RR)$.  The   wavelet coefficients of $f$ are  
\BE \label{eq:wavcoeff}  c_{j,k} = 2^{j} \int_{\mathbf{R}} f(t)  \psi (2^j t - k) dt.  \EE  
In order to determine if a signal can be modelled by a locally bounded function,  one determines the value taken by its { \em uniform H\"older exponent}, denoted by $\Hmin $, and which is defined through a log-log plot regression as  
\BE \label{defhmin} \Hmin = \limsup_{j \to +\infty} \left( \frac{\log \left( \sup_k |c_{j,k}| \right)}{\log (2^{-j})} \right). \EE 
This parameter   is also used for classification,  as shown in Sec.  \ref{sec:phys}. 
 If $\Hmin >0$, then the data can be modelled by a locally bounded function, and  a multifractal analysis based on the pointwise H\"older exponent can be performed. However, as will be shown in Sec. \ref{sec:phys},  for physiological data $\Hmin $ often is found negative. In that case two possible solutions can be chosen: first, one can perform a preprocessing, which consists in a  fractional  integration of the data; alternatively,  if one does not want to alter the nature of the singularities by this smoothing procedure, then, one must pick  a less restrictive  setting for the notion of pointwise singularities which is used.  One possibility is supplied by  $p$-exponents; in this case, the local $L^\infty$ norm in \eqref{eq:reghol} is replaced by a local $L^p$ norm, see \cite{PART1,PART2} for their use in the context of multifractal analysis and \cite{JSW2,GRETSI} for a first use in the context of physiological data. This exponent can be used as soon as data can be modelled by functions  which locally belong to $L^p$.  However, physiological data can prove so irregular that this  is  wrong for all values of $p$, see \cite{JSWPB} for data recorded on marathon runners, and \cite{ACDJ} for  MEG data. In that  case,  one has recourse to the { \em weak scaling exponent}, which is the only pointwise exponent that requires no a priori assumption on the data.  

\subsection{\texorpdfstring{Multifractal formalism and $(\theta,\omega)$-leaders}{Multifractal formalism and leaders}}

\label{sec:math2}

The weak scaling exponent, introduced by Yves Meyer in \cite{MeyWVS}, is required for multifractal analysis when no $p$-exponent can be applied, but one still intends to directly analyze the data, rather than using a regularized version obtained through fractional integration. In fact it is defined in the very general setting of tempered distributions so that in applications,
no  a priori assumption needs to be verified by the data in order to use it. In addition, it has simpler mathematical properties than the other pointwise regularity exponents, see \cite{JSWPB}; let us recall its definition.
Let $f: \RR \rightarrow \RR$ denote a tempered distribution. $f \in C^{s,s'} (x_0)$, if its wavelet coefficients satisfy:
 $\exists C $ $ \forall j,k,  $ $ | c_{j,k} | \leq C 2^{-sj} (1+ | 2^j x_0 -k|)^{-s'}, $ 
 In addition, $ f \in \Gamma^s (x_0)$ if and only if  there exists $s' >0$ such that $f\in C^{s,-s'} (x_0)$. The weak scaling exponent can be  defined as:  $ h^{ws}_f (x_0)= \sup \{ s : \;  f \in \Gamma^s (x_0)\}. $

 Determining $h^{ws}_f$ at each point $x$ is not numerically feasible. Therefore, multifractal analysis provides a global description of the regularity of $f$ in the form of a multifractal spectrum, which describes the size of the set of points where the pointwise regularity exponent takes the same value. These sets of points of the same regularity can have complex geometric structures, and one  describes their size using  fractional dimensions: The  multifractal spectrum is the function that, for each value of regularity exponent $H$ associates the { \em Hausdorff dimensions} of the set of points $x$ such that $h^{ws}_f(x)=H$:
 $$ \dws (H) = \dim_H ( \{ x: h_f^{ws}(x)=H \} ). $$
In practice, multifractal analysis involves estimating $\dws (H)$, which is typically done using  { \em multifractal formalisms}. This approach is, in fact, based on { \em multiresolution quantities  } corresponding to the chosen pointwise regularity exponent. In this study, the data analyzed require the use of  the weak scaling exponent;  this setting requires the the introduction of  { \em $(\theta,\omega)-$leaders } which will be the multiresolution quantities on which  multifractal analysis will be based.  In the following definition,   wavelet coefficients  are still computed using \eqref{eq:wavcoeff},  but the integral has to be understood in the sense of duality between   functions of the Schwartz class and distributions.  

\BD   Let $f$ be a tempered distribution of wavelet coefficients $(c_{j,k})$, and let  $\theta$ and $  \ome $ be two functions   with respectively sub-polynomial and  sub-exponential growth (see \cite{JSWPB,MeyWVS}). The $(\theta, \ome)$-neighbourhood of $(j,k)$, denoted by $V_{ (\theta, \ome)}  (j,k)$  is the set of indices $(j', k')$ satisfying 
 \[ j \leq j' \leq j + \theta (j)  \quad  \mbox{ and } \quad \left| \frac{k}{2^j} -\frac{k'}{2^{j'}} \right| \leq  \frac{\ome (j)}{2^j} . \] 
 The $(\theta, \ome) $-leaders  of $f$ are defined by 
 \BE \label{defomelead} \djk = \sup_{  (j', k')  \in V_{ (\theta, \ome)}  (j,k) } | c_{j',k'} | . \EE   
\ED

Examples of function with sub-polynomial and  sub-exponential growths are supplied by respectively $(\log j))^a$, and $j^a$ for $a >0$, see \cite{JSWPB,ACDJ}, and we will make this choice below.  
For fixed analysis scales $2^{-j}$ the time averages of the  powers of order $p$ of the $\djk$ are referred to as the structure functions: $  S_f (p,j) = 2^{j} \displaystyle\sum_{ k}  | \djk |^p, \ \ \forall p \in \RR. $ These functions exhibit power law behaviors with respect to the analysis scale $2^{-j}$ , in the limit of small scales $2^{-j} \rightarrow 0$: $S_f (p,j) \sim 2^{-j \eta_f(q)} $. More precisely, the { \em scaling function}  $\eta_f(q)$ is  defined as
$$\eta_f (p) =   \displaystyle \liminf_{j \rightarrow + \infty} \;\; \frac{\log \left( S_f (p,j)  \right) }{\log (2^{-j})}.$$ 
Moreover, the Legendre transform of the scaling function
$$ \mathcal{L}(H) =\inf_{ p\in \RR} \left( Hp -\eta_f (p)  +1\right)$$ is called  the {\em Legendre spectrum}; it 
provides an upper bound for the multifractal spectrum: $\forall H$,   $\dws (H) \leq \mathcal{L}(H) $ \cite{JSWPB,ACDJ}. 
 Therefore, for practical applications, the   multifractal spectrum is estimated using by the Legendre spectrum. Multifractal analysis of physiological data has provided a successful alternative to previous wavelet-based methods in several situations where no $p$-exponent could be used, see \cite{JSWPB,ACDJ}.

\subsection{Application on physiological data}

\label{sec:phys} 

The goal of practical multifractal analysis is to estimate the Legendre spectrum. A full estimation of  the function $\mathcal{L}(H) $ is not convenient,  and one rather retains a  few characteristic parameters; these include:
\begin{itemize}
    \item The exponent $H_{min}$, which yields the minimal value taken by the pointwise exponent $h(x)$.
    \item The exponent $c_1^{ws}$, which gives the  value of $h(x)$ mostly met in the data and can be interpreted as a measure of the average smoothness of $f$.
    \item The exponent $c_2^{ws}$, which is related to  the width of the multifractal spectrum and therefore indicates the range of values taken by exponent $h(x)$.
\end{itemize}

\begin{figure}[H]
    \centering
    \includegraphics[scale=0.35]{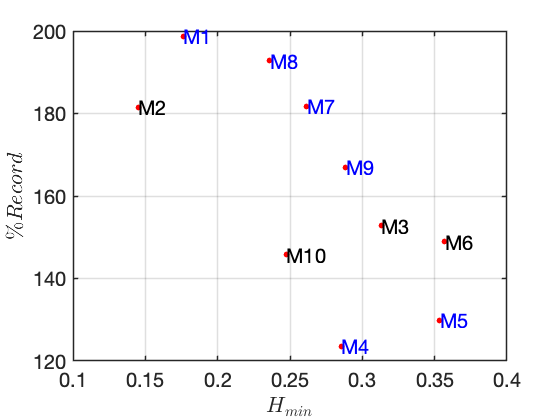}
    \includegraphics[scale=0.35]{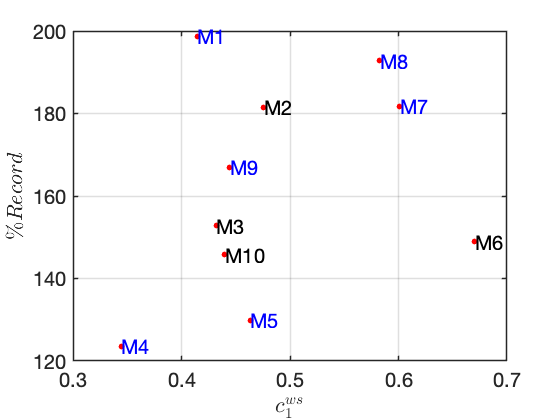}
    \captionsetup{width=0.9\textwidth}
    \caption{Representation of $H_{min}$ (on the left) and $c_1^{ws}$(on the right) as a function of each marathoner's record. Runners in blue participated in the Paris Marathon, while those in black participated in other marathons.  }
    \label{hmin-c1-record}
\end{figure}

\begin{figure}[H]
    \centering
    \includegraphics[scale=0.35]{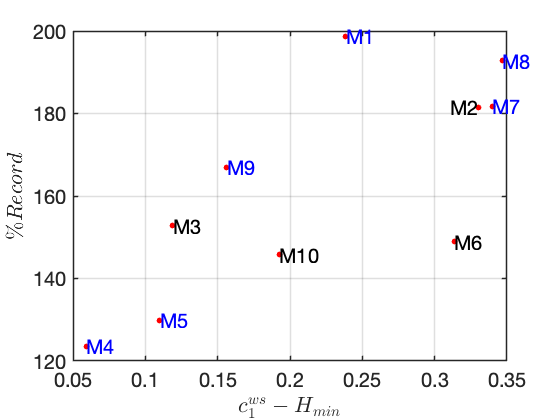}
    \includegraphics[scale=0.35]{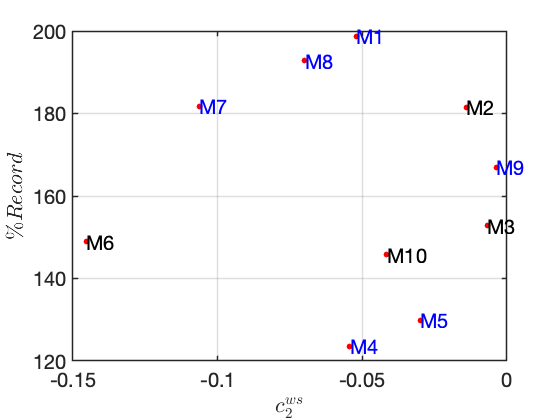}
    \captionsetup{width=0.9\textwidth}
    \caption{Representation of $c_1^{ws}-H_{min}$ (on the left) and $c_1^{ws}$(on the right) as a function of each marathoner's record. Runners in blue participated in the Paris Marathon, while those in black participated in other marathons.  }
    \label{c1-hmin-c2-record}
\end{figure}

In this work, these  multifractality parameters are estimated using log-log regressions  of quantities based on $(\theta,\omega)-$leaders), the { \em log-cumulants} see  \cite{wendt2007multifractality,JSWPB}. 
Thus, we present on Table \ref{tab-anal} the three multifractal parameters linked to the multifractal spectrum, based on $(\theta,\omega)-$leaders, of 10 heart rates signals (in beats per minute).  In addition, we compare these parameters in Figure \ref{hmin-c1-record} and \ref{c1-hmin-c2-record}.
\begin{table}[H]
\begin{center}
\resizebox{170mm}{!}{
 \begin{tabular}{||c||*{11}{m{1.95cm}|}|} 
 \hline
  & \textcolor{blue}{M1} & M2 & M3 & \textcolor{blue}{M4} & \textcolor{blue}{M5} & M6 \textbf{Women} & \textcolor{blue}{M7} & \textcolor{blue}{M8} & \textcolor{blue}{M9} & M10 \\ [0.5ex] 
 \hline\hline
 $H_{min}$& \textcolor{blue}{0.1761} & 0.145 & 0.3131 & \textcolor{blue}{0.2854} & \textcolor{blue}{0.3554} & 0.357 & \textcolor{blue}{0.2615} & \textcolor{blue}{0.2361} & \textcolor{blue}{0.2884} & 0.2474 \\ 
\hline
$c_1^{ws}$& \textcolor{blue}{0.4145} & 0.4756 & 0.4321 & \textcolor{blue}{0.3444} & \textcolor{blue}{0.4635} & 0.6704 & \textcolor{blue}{0.6013} & \textcolor{blue}{0.5827} & \textcolor{blue}{0.4447} & 0.4399 \\
\hline
 $c_2^{ws}$& \textcolor{blue}{-0.0521} & -0.0141 & -0.0068 & \textcolor{blue}{-0.0543} & \textcolor{blue}{-0.03} & -0.1455 & \textcolor{blue}{-0.1063} & \textcolor{blue}{-0.0699} & \textcolor{blue}{-0.0038} & 0.4399 \\
 \hline
 \end{tabular}}
\end{center}
\captionsetup{width=0.9\textwidth}
\caption{Representation of different multifractal parameters $(H_{min},c_1^{ws},c_2^{ws})$ of the Heart rate of each marathon runner.}
\label{tab-anal}
\end{table}

\subsection{Results and Discussion}
The findings suggest that multifractal analysis can provide valuable feedback for optimizing pacing strategies. Indeed, the analysis revealed that multifractal parameters could detect changes in physiological signals due to fatigue, especially around the 30th kilometer mark. The study also shows that better-ranked runners had more uniform regularity in their physiological signals.
Figure \ref{hmin-c1-record} shows that the better the ranking, the larger the $H_{min}$ value. Conversely, when examining the figure for $c_1^{ws}$, which represents the regularity almost everywhere versus the record, the opposite trend is observed: marathoners with lower performance levels have a higher $c_1^{ws}$  value compared to those at the top ranks. These two observations suggest that the better the ranking, the narrower the marathoner's multifractal spectrum. In other words, higher-performing runners exhibit more uniform regularity, indicating less "fractality", i.e. the regularity exponent $h(x)$ varies over a small interval. 
Besides,  Figure \ref{c1-hmin-c2-record}, particularly the $c_1^{ws}-H_{min}$ graph, confirms our conclusion: the smaller the gap, the better the marathoner's ranking. When discussing the parameter $c_2^{ws}$, we focus on its influence on the spectrum's concavity. As $c_2^{ws}$ approaches zero, the spectrum becomes less concave and narrower, indicating a more monofractal nature, where the pointwise regularity parameter remains constant. Conversely, as $c_2^{ws}$ moves away from zero in the negative direction, the spectrum becomes more concave and broader, thus indicating higher "multifractality". The analysis of the $c_2^{ws}$ graph in relation to the record supports our conclusions: athletes with lower performance exhibit more negative $c_2^{ws}$ values, corresponding to a higher degree of "multifractality".

\begin{figure}[H]
    \centering
    \includegraphics[scale=0.5]{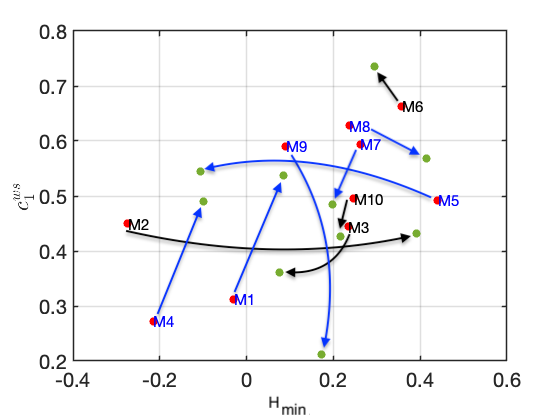}
    \captionsetup{width=0.9\textwidth}
    \caption{Evolution of the couple $(H_{min},c_1^{ws})$ between the first half (red dots) and the last fourth (green dots) of the marathon. Runners in blue participated in the Paris Marathon, while those in black participated in other marathons. }
    \label{hmin-c1-part}
\end{figure}

We enhanced the previous analysis by examining variations in multifractal parameters throughout the marathon. It is well established that during the final 12 kilometers, many runners experience a significant increase in difficulty, with Borg RPE (Rate of Perceived Exertion) scores exceeding 15/20, indicating strenuous effort. It is intriguing to explore how these changes impact  multifractality parameters. Figure. \ref{hmin-c1-part} illustrates the  evolution of the multifractal parameters $H_{min}$ and $c_1^{ws}$ between the first and last quarters of the marathon,
emphasizing the differences in physiological responses to fatigue beyond the 30th kilometer. Note that these two parameters do not necessarily exhibit the same type of variation:  the variations of $H_{min}$ and $c_1^{ws}$ for marathoner M1, who finished last, and M4, who finished first, are very similar. This indicates that both the first and the last runners experienced the same variation in spectrum, meaning the same change in regularity. Despite finishing last, M1 managed to adjust and pace his race similarly to M4. It means that, despite his lower performance level, he was able to adequately self-pace his race, which is an important factor of performance (in addition to the maximal oxygen uptake (VO2max), the endurance i.e. the ability for sustaining a high fraction of VO2max throughout the marathon, and the running economy i.e. the stride efficiency). 

The factors influencing marathon performance, as outlined by Michael Joyner, a prominent exercise physiologist, are based on physiological and biomechanical aspects that determine a runner's potential. In the 1990s, Joyner put forth a theoretical model that postulated that marathon performance could be optimized by focusing on three primary factors.
The maximal oxygen uptake (VO2 Max) represents the maximum rate at which an individual can consume oxygen during intense exercise. It is indicative of the aerobic capacity and is a pivotal determinant of endurance performance. Its impact  on endurance performance is significant: a higher VO2 max enables a runner to maintain a faster pace over longer distances. The typical range for VO2 max in elite marathoners is 70 for female and 85 ml.kg-1.min-1. The present subjects have a VO2max ranged between 44 and 60 ml.kg-1.min-1 that can explain that they are not in the elite marathon runner category (even considering their age and gender). 
The second marathon performance factor is the Lactate Threshold (LT). The term "lactate threshold" is used to describe the exercise intensity at which lactate begins to accumulate in the blood. It serves as an indicator of the efficiency with which the body is able to clear lactate, a byproduct of anaerobic metabolism.  The impact of this phenomenon is the following: runners pace  closer to their lactate threshold allows for a faster running speed without the accumulation of elevated levels of lactate, which can otherwise lead to fatigue. An enhanced LT enables to sustain a greater proportion of  VO2 max for extended durations.
The third factor of performance level is the Running Economy (RE). It  is defined as the efficiency with which a runner utilizes oxygen at a given pace. It is indicative of the energy expenditure associated with running at a given velocity. An enhanced running economy enables the runner to expend less energy at a given pace, thereby facilitating the maintenance of faster speeds over the marathon distance. Factors such as biomechanics, muscle fiber composition, and even psychological factors can influence running economy.

\subsection{Additional Considerations}

While Joyner's model puts particular emphasis on these three physiological factors, he and other researchers have also identified the importance of:
\begin{itemize}
    \item \textbf{Glycogen Stores:} The capacity to store and utilize glycogen in an efficient manner is of paramount importance for maintaining optimal energy levels throughout the marathon.
    \item \textbf{Mental Toughness:} Psychological resilience and the capacity to manage pain and discomfort can markedly impact performance.
    \item \textbf{Heat Tolerance:} Running in warmer conditions necessitates superior thermoregulation and hydration strategies to maintain performance.
\end{itemize}
Joyner's contributions to this field established a conceptual framework for understanding the limits of human endurance performance. His model has been employed to investigate the theoretical potential for marathon times under optimal conditions. Nevertheless, this approach did not consider the optimization of pacing strategy, as the constant speed was assumed to be  optimal. However, a recent study  analyzing the best performance on marathon, showed that marathon performance depends on pacing oscillations between  asymmetric extreme values \cite{pycke2022marathon}.

\begin{figure}[H]
    \centering
    \includegraphics[scale=0.28]{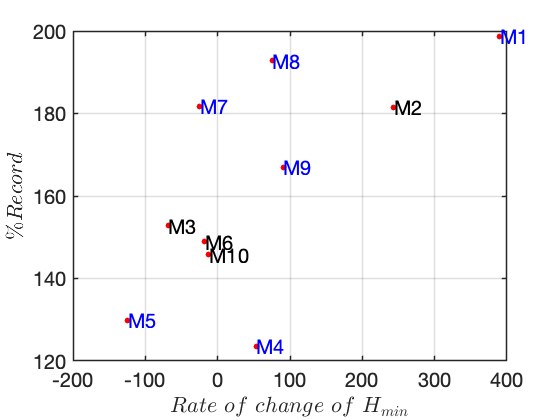} 
    \includegraphics[scale=0.28]{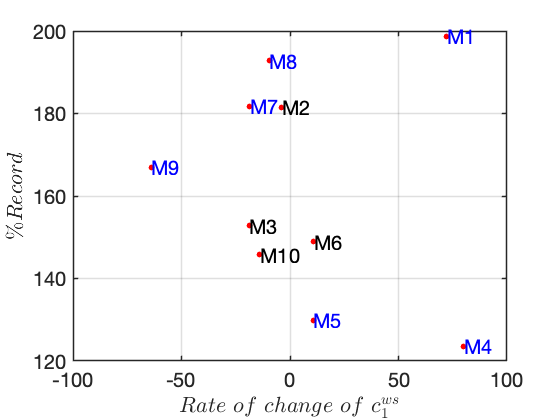} 
    \includegraphics[scale=0.28]{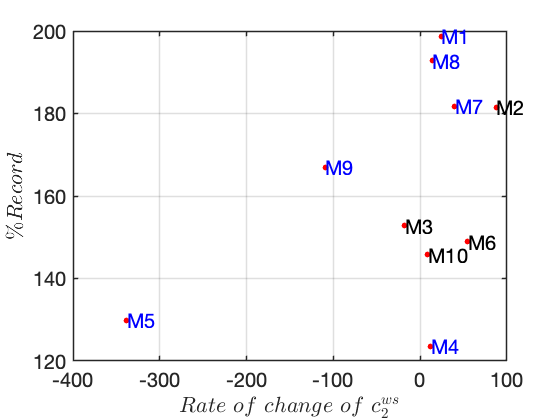}
    \captionsetup{width=0.9\textwidth}
    \caption{Representation of the rate of change of $H_{min}$ and $c_1^{ws}$  as a function of each marathoner's record. Runners in blue participated in the Paris Marathon, while those in black participated in other marathons. }
    \label{evol-c1-hmin}
\end{figure}

The  variation of parameters between the two parts of the race provides insight into the evolution of the corresponding quantity. However, to assess the significance of this change, it must be compared to its initial value. Therefore, in the perspective of self-improving  the runner's performance in the next marathon, we propose  the examination of the rate of change of the multifractal parameters $(H_{min}$, $c_1^{ws}$and $c_2^{ws})$ in Figure \ref{evol-c1-hmin},  as a biofeedback for improving the pace management that could constitute the fourth dimension of the marathon performance. In light of the necessity for self-improvement in order to enhance performance in the forthcoming marathon, we put forth the proposition that the examination of the rate of change of  $(H_{min}$, $c_1^{ws}$ and $c_2^{ws})$ could serve as a biofeedback mechanism for the improvement of pace management. This could be proposed as the fourth dimension of the marathon performance. 

\section{Conclusion and limitations}

In this study, an advanced technique, the  wavelet-based multifractal formalism,  has been employed to quantify the multifractality of time series collected during marathons. 
This technique permits the classification and modelling of physiological data using parameters based on scale invariance exponents. Indeed, multifractal analysis provides a more comprehensive view of the individual performance of marathon runners, identifying alterations in physiological signals resulting from fatigue and optimizing race pace strategies. Therefore, multifractal analysis offers a comprehensive perspective on individual performance offering a new way to understand the complex dynamics of physiological signals during marathons. This approach can help optimize training and pacing strategies to improve performance.


We examined the multifractality characteristics of marathons achieved by a diverse group of recreational runners, including individuals of varying age, gender, and performance levels. Considering the aforementioned diversity, we standardized the performance in percentage of the world performance for each individual category. Despite this standardization,  a more comprehensive data set would be  necessary in order to confirm our conclusions, encompassing not only the heart rate but also the cadence and speed, and bearing on more runners in order to apply statistical tools. It is imperative to persuade a greater number of marathon runners to share their data with the objective of validating it in open source. This would facilitate the development of an algorithm capable of providing an index of race optimization, thereby preventing the majority of recreational marathoners, and even some elites, from encountering the phenomenon commonly referred to as the "marathon wall." In the context of the Olympic Games, the allocation of medals is contingent upon pacing strategies, underscoring the importance of this endeavor.

\section{Appendix} 

In this section, we show how to perform  the reconnection procedure  so that it does not add spurious  pointwise singularities  as long as the pointwise H\"older exponent present in the data are below $H= 1$. Recall that this procedure consists in eliminating a part of the signal where data are altered and  reconnecting the two portions on the left and right of the eliminated interval $[x_0, y_0]$.

We first consider the simple case of a continuous signal. In that case, it is natural to pick for pointwise regularity exponent the H\"older exponent. The reconnection is continuous if the points $x_0$ and $y_0$ are picked such that $f(x_0) = f(y_0)$, which we assume. 
Let $\alpha \in (0,1)$ be such that $\alpha <\min ( h^p_f (x_0) , \; h^p_f (y_0))$.  The initial function $f$ satisfies : $f\in C^{\alpha} (x_0) $ and $f\in C^{\alpha} (y_0) $. The reconnected function is  the function $g$ defined by 
\[ \left\{ \begin{array}{rll}
    g(x)  & = f(x)  &  \mbox{ if }  x \leq x_0 \\
     & = f(x+ y_0 -x_0) & \mbox{ if }  x  > x_0 . 
\end{array} \right. \] 
Let us check that the reconnected function $g$ satisfies $g\in C^{\alpha} (x_0) $, i.e. that 
\begin{equation} \label{calph} 
    | g(x) -g(x_0)| \leq C | x-x_0|^\alpha . 
\end{equation} 
Since $g$ coincides with $f$ for $x < x_0$  \eqref{calph}  holds in that case. 
Assume now that $x > x_0$. Then  
\[  | g(x) -g(x_0)|= |f(x+ y_0 -x_0)-f(y_0) |  \leq C | x-x_0|^\alpha .  \] 
In conclusion, the regularity of the reconnected function is larger than the  lowest regularity  at the two initial points.

We now consider the case of functions that are not continuous. We assume that $f \in L^p_{loc}$, in which case, it is natural to pick for pointwise regularity exponent the $p$-exponent; in this setting the notion of continuous reconnection does not make sense any more; however, this does not mean that all reconnections lead to the same regularity: consider for instance  the case where $f$ is $C^1$ function  and one operates a discontinuous reconnection. Then the $p$-exponent at the reconnection will vanish  and will thus be lower than the $p$-exponent  at the initial points (which are larger than 1). 
In order to determine the procedure to make a smooth reconnection in the $L^p$  case, we first notice  that we can eliminate the case where $h^p_f (x_0) <0$ or $h^p_f (y_0) <0$: indeed, in that case  let us assume e.g. that $h^p_f (x_0) <0$ and $h^p_f (x_0) <h^p_f (y_0)$. It follows that the Taylor polynomial at $x_0 $  vanishes, so that, for $r$ small enough, 
\[ \forall   \alpha < h^p_f (y_0), \qquad \int_{ x_0 -r}^{x_0 + r} | f(x) |^p dx \leq C r^{\alpha p +1 }. \]  Since  $h^p_f (x_0) <h^p_f (y_0)$, we also have 
\[   \int_{ y_0 -r}^{y_0 + r} | f(x) |^p  dx \leq C r^{\alpha p +1 }, \] 
and a simple reconnection can be performed and yields  
\[  \int_{ x_0 -r}^{x_0 + r} | g(x) |^p dx  \leq C r^{\alpha p +1 }; \] 
in that case any reconnection has the property of not lowering  the $p$-exponent.  
There remains to consider the case where 
\[ 0 <  \min ( h^p_f (x_0) , \; h^p_f (y_0))  <1 . \] 
We can assume that this minimum is $h^p_f (x_0)$.  In that case, the Taylor polynomial at $x_0 $  and $y_0$  take  constant values, respectively denoted by  $C_0$ and $D_0$. 
Recall that, since the $p$-exponent is positive at $x_0$ and $y_0$, it follows that they are { \em Lebesgue points } of  $f$,
i.e. the limits 
\[  \lim_{r \rightarrow 0} \frac{2}{r}  \int_{ x_0 -r}^{x_0 + r}  f(x)  dx  \quad \mbox{ and } \quad  \lim_{r \rightarrow 0} \frac{2}{r}  \int_{ y_0 -r}^{y_0 + r}  f(x)  dx  \] exist and  are  called  { \em Lebesgue values} of $f$ at $x_0$ and $y_0$, and furthermore $C_0$  and $D_0$  respectively coincide with these limits, see \cite{JaffMel,PART1}. 
Let $ \alpha < \min ( h^p_f (y_0), h^p_f (y_0))$;  then, 
for $r$ small enough, 
\[ \int_{ x_0 -r}^{x_0 + r} | f(x) -C_0 |^p dx \leq C r^{\alpha p +1 }\quad \mbox{ and } \quad   \int_{ y_0 -r}^{y_0 + r} | f(x)-C_0  |^p \leq C r^{\alpha p +1 }, \] 
so that \[  \int_{ x_0 -r}^{x_0 + r} | g(x) -C_0 |^p  dx \leq C r^{\alpha p +1 };  \] 
additionally, one easily checks that, if the Lebesgue values at the reconnection points $x_0$ and $y_0$ differ, then the $p$-exponent of $g$ at $x_0$ vanishes, thus creating a spurious singularity.  
In conclusion, in order not to create  artificial singularities, the reconnection has to be done at points where the Lebesgue values of $f$ coincide.  Note that, if one disposes of a wavelet decomposition of $f$, then the Lebesgue value of $f$ at $x_0$ is also given   by the limit of the partial sums of the wavelet series that allows to reconstruct $f$, i.e. by 
\[ C_0 = \lim_{J \rightarrow + \infty} \sum_{j \leq J} \sum_k C_{j,k} \psi_{j,k} (x) ; \] 
it follows that, in practice,  the equality of the Lebesgue values at $x_0$ and $y_0$  can be checked using the   wavelet expansion of $f$.

\bibliography{biblio}

\begin{thebibliography}{10}

\bibitem{abry2009multifractal}
P.~Abry, P.~Chainais, L.~Coutin, and V.~Pipiras.
\newblock Multifractal random walks as fractional wiener integrals.
\newblock {\em IEEE Transactions on Information Theory}, 55(8):3825--3846,
  2009.

\bibitem{ACDJ}
P.~Abry, P.~Ciuciu, M.~Dumeur, S.~Jaffard, and G.~Sa\"es.
\newblock Multifractal analysis based on the weak scaling exponent and
  applications to meg recordings in neuroscience.
\newblock {\em Preprint}, 2024.

\bibitem{CRAS2019}
P.~Abry, H.~Wendt, S.~Jaffard, and G.~Didier.
\newblock Multivariate scale-free temporal dynamics: From spectral ({F}ourier)
  to fractal (wavelet) analysis.
\newblock {\em Comptes Rendus de l'Acad\'emie des Sciences}, 20(5):489--501,
  2019.

\bibitem{ariza2019joint}
A.~B Ariza-Villaverde, P.~Pav{\'o}n-Dom{\'\i}nguez, R.~Carmona-Cabezas, E.~G.
  de~Rav{\'e}, and F.~J. Jim{\'e}nez-Hornero.
\newblock Joint multifractal analysis of air temperature, relative humidity and
  reference evapotranspiration in the middle zone of the guadalquivir river
  valley.
\newblock {\em Agricultural and Forest Meteorology}, 278:107657, 2019.

\bibitem{Bacry2001}
E.~Bacry, J.~Delour, and J.F. Muzy.
\newblock Multifractal random walk.
\newblock {\em Phys. Rev. E}, 64(2):026103, 2001.

\bibitem{JSWPB}
W.~Ben~Nasr, V.~Billat, S.~Jaffard, F.~Palacin, and G.~Sa\"es.
\newblock The weak scaling multifractal spectrum: Mathematical setting and
  applications to marathon runners physiological data.
\newblock {\em To appear in the proceedings of the FARF IV conference}, 2024.

\bibitem{billat2009detection}
V.~Billat, L.~Mille-Hamard, Y.~Meyer, and E.~Wesfreid.
\newblock Detection of changes in the fractal scaling of heart rate and speed
  in a marathon race.
\newblock {\em Physica A: Statistical Mechanics and its Applications},
  388(18):3798--3808, 2009.

\bibitem{Billat19}
V.L. Billat, F.~Palacin, M.~Correa, and J.R Pycke.
\newblock Pacing strategy affects the sub-elite marathoner's cardiac drift and
  performance.
\newblock {\em Front Psychol}, 10:3026, 2020.

\bibitem{gerasimova2014wavelet}
E.~Gerasimova, B.~Audit, S-G. Roux, A.~Khalil, O.~Gileva, F.~Argoul,
  O.~Naimark, and A.~Arneodo.
\newblock Wavelet-based multifractal analysis of dynamic infrared thermograms
  to assist in early breast cancer diagnosis.
\newblock {\em Frontiers in physiology}, 5:176, 2014.

\bibitem{huynh2023multifractality}
P.~K. Huynh, D.~Nguyen, G.~Binder, S.~Ambardar, T.~Q. Le, and D.~V. Voronine.
\newblock Multifractality in surface potential for cancer diagnosis.
\newblock {\em The Journal of Physical Chemistry B}, 127(31):6867--6877, 2023.

\bibitem{ivanov1999}
P.C. Ivanov, L.A.~Nunes Amaral, A.L. Goldberger, S.~Havlin, M.G. Rosenblum,
  Z.R. Struzik, and H.E. Stanley.
\newblock Multifractality in human heartbeat dynamics.
\newblock {\em Nature}, 399:461--465, 1999.

\bibitem{Jaffard2004}
S.~Jaffard.
\newblock Wavelet techniques in multifractal analysis.
\newblock In M.~Lapidus and M.~van Frankenhuijsen, editors, {\em Fractal
  Geometry and Applications: A Jubilee of Beno\^{\i}t Mandelbrot, Proc. Symp.
  Pure Math.}, volume 72(2), pages 91--152. AMS, 2004.

\bibitem{JaffMel}
S.~Jaffard and C.~Melot.
\newblock Wavelet analysis of fractal boundaries.
\newblock {\em Communications In Mathematical Physics}, 258(3):513--565, 2005.

\bibitem{PART1}
S.~Jaffard, C.~Melot, R.~Leonarduzzi, H.~Wendt, S.~G. Roux, M.~E. Torres, and
  P.~Abry.
\newblock p-exponent and p-leaders, {P}art {I}: {N}egative pointwise
  regularity.
\newblock {\em Physica A}, 448:300--318, 2016.

\bibitem{GRETSI}
S.~Jaffard, G.~Saes, W.~Ben~Nasr, F.~Palacin, and V.~Billat.
\newblock Analyse multifractale des donn\'ees physiologiques de marathoniens.
\newblock In {\em Colloque sur le Traitement du Signal et des Images. GRETSI
  2022}, 2022.

\bibitem{JSW2}
S.~Jaffard, G.~Sa\"es, W.~Ben~Nasr, F.~Palacin, and V.~Billat.
\newblock A review of univariate and multivariate multifractal analysis
  illustrated by the analysis of marathon runners physiological data.
\newblock In {\em Mathematical Analysis and Applications, Plenary Lectures,
  ISAAC 2021}. Springer Proceedings in Mathematics and Statistics, 2023.

\bibitem{jianga2018multifractal}
Z.-Q. Jiang, W.-J. Xie, W.-X. Zhou, and D~Sornette.
\newblock Multifractal analysis of financial markets.
\newblock {\em Research Center for Econophysics, East China University of
  Science and Technology}, 82(12):1--145, 2018.

\bibitem{kantelhardt2002multifractal}
J.~W. Kantelhardt, S.~A. Zschiegner, E.~Koscielny-Bunde, S.~Havlin, A.~Bunde,
  and H.~E. Stanley.
\newblock Multifractal detrended fluctuation analysis of nonstationary time
  series.
\newblock {\em Physica A}, 316(1):87--114, 2002.

\bibitem{PART2}
R.~Leonarduzzi, H.~Wendt, S.~G. Roux, M.~E. Torres, C.~Melot, S.~Jaffard, and
  P.~Abry.
\newblock p-exponent and p-leaders, {P}art {II}: {M}ultifractal analysis.
  {R}elations to {D}etrended {F}luctuation {A}nalysis.
\newblock {\em Physica A}, 448:319--339, 2016.

\bibitem{MeyWVS}
Y.~Meyer.
\newblock {\em Wavelets, vibrations and scalings}.
\newblock CRM Ser. AMS Vol. 9,, Presses de l'Universit\'e de Montr\'eal, Paris,
  1998.

\bibitem{muzy1991wavelets}
J.-F. Muzy, E.~Bacry, and A.~Arneodo.
\newblock Wavelets and multifractal formalism for singular signals: Application
  to turbulence data.
\newblock {\em Physical review letters}, 67(25):3515, 1991.

\bibitem{pycke2022marathon}
JR. Pycke and V.~Billat.
\newblock Marathon performance depends on pacing oscillations between non
  symmetric extreme values.
\newblock {\em International Journal of Environmental Research and Public
  Health}, 19(4):2463, 2022.

\bibitem{wendt2007multifractality}
H.~Wendt and P.~Abry.
\newblock Multifractality tests using bootstrapped wavelet leaders.
\newblock {\em IEEE Transactions on Signal Processing}, 55(10):4811--4820,
  2007.

\bibitem{wesfreid05a}
E.~Wesfreid, V.~Billat, and Y.~Meyer.
\newblock Multifractal analysis of heartbeat time series in human races.
\newblock {\em Appl. Comput. Harmon. Anal.}, 2010.

\end{thebibliography}

\end{document}